\documentclass{jpconf}

\usepackage{graphicx}
\usepackage[numbers]{natbib}
\usepackage{amssymb}
\usepackage[rightcaption]{sidecap}

\begin{document}

\title{Interstellar Dust and Gas in the Heliosphere}

\author{Jonathan Slavin}

\address{Center for Astrophysics $\mid$ Harvard \& Smithsonian, 60 Garden Street, Cambridge, MA 02138}

\ead{jslavin@cfa.harvard.edu}

\begin{abstract}
Interstellar dust and gas that enter the heliosphere provide us with important clues about both the heliosphere and the local interstellar medium (LISM). The picture we have of the Local Interstellar Cloud (LIC) from both \emph{in situ} detections and absorption line data presents questions that have important implications for the origins and evolution of the cloud. New detections of $^{60}$Fe on Earth in deep sea crusts and Antarctic snow cores provide evidence for the role of supernovae in shaping the LISM. We discuss our models for the evolution of the LIC inside the Local Bubble and possible explanations for the source of the supernova produced dust.
\end{abstract}

\section{Introduction}
The interstellar dust and gas that flow into the heliosphere are coming from the interstellar cloud that surrounds the Solar System known as the Local Interstellar Cloud (LIC). To understand both the LIC and the heliosphere we need information on the state of the LISM. In turn, a knowledge of the current state of the LISM can provide important information on the history and evolution of the LISM, in particular the LIC and other nearby clouds (the Complex of Interstellar Clouds or CLIC) and the Local Bubble that surrounds the clouds. The detailed observations that we have of the LISM can then help us to understand interstellar processes and evolution more broadly.

To focus our discussion of the nature of the nearby ISM, I have chosen to examine four outstanding questions regarding the LISM. The first two are longstanding questions while the second two are relatively new:
\begin{itemize}
    \item How did the low density clouds of the CLIC get into the Local Bubble?
    \item Why is the grain size distribution of interstellar dust in the heliosphere so strange?
    \item What is the source of the $^{60}$Fe peaks seen in deep sea crusts?
    \item What does the $^{60}$Fe in Antarctic snow cores tell us?
\end{itemize}
These questions have some interesting interconnections and could hold important clues to how the interstellar dust and gas evolve.

\section{The origins of the CLIC}\label{sect:LISM}
To understand why the existence of the CLIC within the Local Bubble is mysterious, one needs to be aware of the way the thermal phases in the ISM are created and evolve, as inferred from both observations and theoretical modeling. It has been found that the interstellar plasma tends to roughly segregate into different temperature-density-ionization phases. These are cold ($T \sim 100$ K, $n \sim 50$ cm$^{-3}$), warm ($T \sim 6000 - 9000$ K, $n \sim 0.2$ cm$^{-3}$) and hot ($T \sim 10^6$ K, $n \sim 5\times10^{-3}$ cm$^{-3}$). The warm and cold phases can be in approximate thermal equilibrium given known heating and cooling processes whereas the hot phase is associated with shock heating and is not in equilibrium but cools slowly. The warm phase is seen to be widely and fairly smoothly distributed \citep[see e.g., H$\alpha$ maps,][]{Haffner_etal_2003}, whereas the cold phase gas tends to be confined to clouds. Such structures are also seen in Galaxy scale simulations with realistic heating and cooling rates in which supernovae drive the medium \citep[e.g.,][]{Hill_etal_2018}.

The Local Bubble, which surrounds the CLIC, is filled with hot gas \citep{Mccammon_etal_1983,Galeazzi_etal_2014,Liu_etal_2017} and must therefore have been created by fast shocks. In fact because of the size and thermal pressure of the bubble it must have bee created by multiple supernovae \citep[e.g.,][]{Breitschwerdt_etal_2016}. It is well known \citep{McKee+Ostriker_1977,Slavin_etal_2017} that supernova remnant (SNR) shocks propagating through a cloudy medium will sweep around and leave behind dense clouds that they encounter. However, that implies that the CLIC must have been denser than the surrounding medium. That would seem to contradict the fact that the clouds now have densities ($n \sim 0.2$ cm$^{-3}$) that are characteristic of the lower density, warm inter-cloud medium.

The solution to this problem is that the clouds \emph{were} dense and cold at the time they were encountered in the ISM by the first shock that carved out the Local Bubble. We must then ask how (and if) the clouds could have evolved from that cold, dense state to the warm, low density state that we see them in now. This is a complex question since it involves a number of processes including the heating and cooling rates caused by photoionization heating and radiative cooling as well as the effects of the shock(s) that hit the clouds and shear flow as the surrounding gas sweeps past the clouds.

\begin{figure}[ht!]
    \centering
    \includegraphics[width=0.8\textwidth]{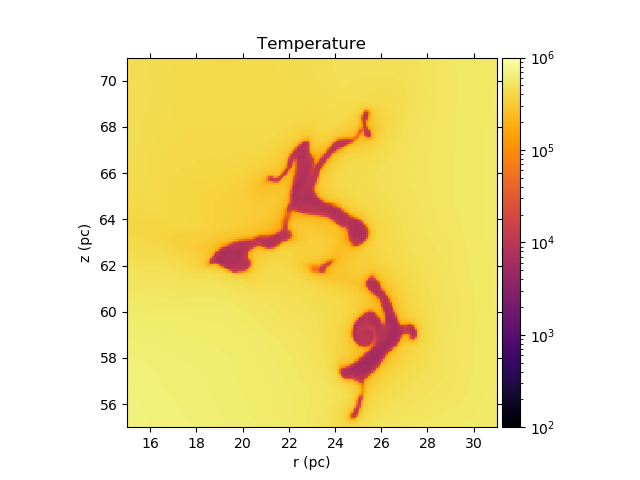}
    \caption{Temperature for a complex of clouds embedded in the hot bubble in our magnetohydrodynamical model for the LISM. The model has two supernova explosions that occur at an interval of $5\times10^5$ yr. The clouds, which were initially cold and dense have rebounded after having been shocked and have warmed to temperatures similar to those derived for the CLIC.}
    \label{fig:CLIC_temp}
\end{figure}

To assess the possibility that the clouds could be heated in this way to achieve their current state, we have carried out a series of numerical simulations. The heating and cooling rates used were realistic and allow for two stable temperature equilibria to exist, one for cold clouds and one for warm intercloud gas. A magnetic field is used with a strength typical of that derived for the ISM (5 $\mu$G) and a supernova explosion is set off near a complex of cold clouds. We have experimented with the parameters of the simulations, setting off a second explosion and sometimes a third at intervals of $\sim 5\times10^5$ yr. A general finding is that two or more explosions are necessary. We find that indeed, after being compressed by the first shocks, the clouds do rebound and at late times as the bubble pressure drops near the currently inferred values of $P/k_B \sim 7000$ cm$^{-3}\,$K, the clouds are mostly warm with temperatures near those observed for the CLIC (see Figure \ref{fig:CLIC_temp}). (More details of the simulations and results will be presented in a forthcoming paper.) Thus we conclude that the LIC was originally cold and dense and has been subject to at least two shocks.

\section{The interstellar grain size distribution in the heliosphere}
\begin{figure}[ht!]
    \centering
    \includegraphics[width=0.5\textwidth]{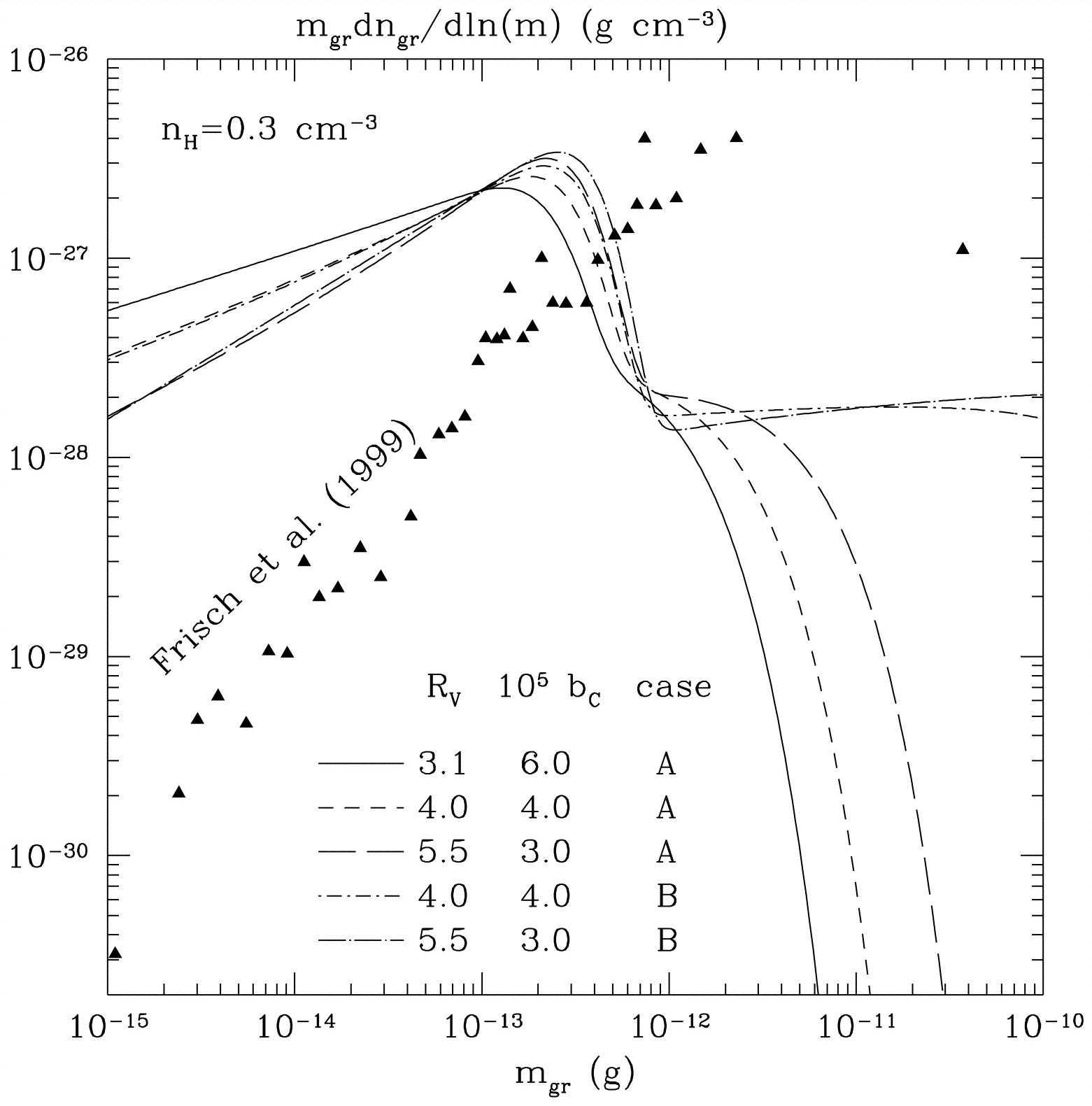}
    \caption{Comparison of grain size distribution for interstellar grains observed in the heliosphere with size distribution models derived from observations \citep[from][used with permission]{Weingartner+Draine_2001}. The heliospheric distribution (triangles) is clearly extremely discrepant.}
    \label{fig:size_distn}
\end{figure}

Another longstanding problem is the grain size distribution detected by several spacecraft in the heliosphere, especially Ulysses \citep{Frisch_etal_1999,Krueger_etal_2015A}. As pointed out by Frisch et al. \citep{Frisch_etal_1999}, Weingartner \& Draine \citep{Weingartner+Draine_2001}, and Draine \citep{Draine_2009} the detected grain size distribution is far from that inferred for the ISM and is in fact not consistent with that distribution (see Figure \ref{fig:size_distn}). Since the charged grains in the LIC will be filtered at the heliopause, the fact that the distribution lacks small grains is not surprising. However, the large excess of large grains is. Such large grains, if they were prevalent in the ISM, would produce essentially wavelength independent (gray) extinction to an extent that would be inconsistent with the wavelength dependence of the many, many extinction curves observed for Galactic sources. So what has caused this odd size distribution? Since shocks tend to erode large grains rather than enhance them, that part of our proposed history for the LIC only exacerbates the problem. We discuss a possible solution to this mystery below (\S\ref{sect:capture}).

\section{$^{60}$Fe in deep sea crusts}
\begin{figure}[ht!]
    \centering
    \includegraphics[width=0.7\textwidth]{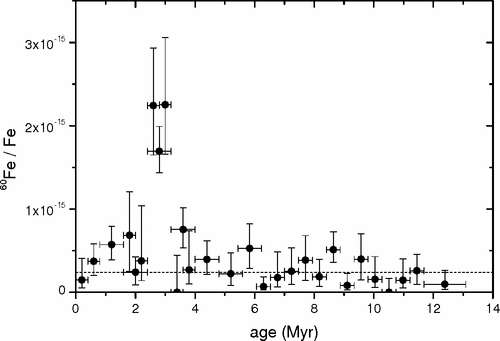}
    \caption{$^{60}$Fe/Fe ratios versus the age of the crust layer \citep[From][used with permission]{Knie_etal_2004}. The peak at $\sim 2.8$ Myr is evidence for an influx of supernova-produced dust reaching the Earth at that time.}
    \label{fig:crust_60Fe}
\end{figure}

A different piece of evidence that may bear on the history of the LISM is the discovery of a peak in the $^{60}$Fe isotope fraction in deep sea ferromanganese crusts \citep{Knie_etal_2004,Fitoussi_etal_2008}. As can be seen in Figure \ref{fig:crust_60Fe}, there was a peak at about 2.8 Myr ago. $^{60}$Fe is not produced in abundance by cosmic rays, but is produced in core collapse supernova (SN) explosions. Because Fe is ionized essentially everywhere in the ISM except inside of very dense clouds, in order for the $^{60}$Fe, to be transported to the Earth's surface through the heliosphere, it had to have been in the form of dust. There is abundant evidence that core collapse SNe produce dust and so the $^{60}$Fe spike can be interpreted as evidence of a nearby SN that exploded roughly $2\times10^6$ yr ago and created dust that was transported to the Earth.  Because the half life of $^{60}$Fe is 2.6 Myr, the SN could not have exploded much earlier than that.

We note that the Earth was not in the LIC at that time so the connection to the state of the LIC in particular is not clear. However, it is quite likely that the explosion that created the dust was one of the explosions that created the Local Bubble, as has been modeled by \citep{Breitschwerdt_etal_2016}. The LIC was also likely to have been affected by the shock generated by that explosion.

\section{$^{60}$Fe in Antarctic snow cores}
More directly relevant to the LIC is the detection of $^{60}$Fe in Antarctic snow cores as reported by \citep{Koll_etal_2019}. Those authors eliminated terrestrial and cosmogenic sources via the $^{60}$Fe/$^{53}$Mn ratio and conclude that the $^{60}$Fe is of interstellar origin. These cores include snow that has precipitated over the past 20 yr, so this is evidence that SN created dust is present in the LIC now. However, we do not know when (and how) the dust got into the LIC. It seems likely that that dust was also generated in a SN that helped to form the Local Bubble, however how the dust was transported and captured by the cloud is uncertain.

\section{Source and transport of $^{60}$Fe dust}
Aiming to understand the source of the $^{60}$Fe in deep sea crusts, Schulreich et al. \citep{Schulreich_etal_2017} modeled the dust transport from a SN that contributed to the heating of the Local Bubble. For their model of Local Bubble evolution, they use 16 SNe going off over time. They tied their model to the locations and history of nearby OB associations which could plausibly (based on their stellar content) have produced SNe during the relevant time period to produce the dust. They assume that the dust moves with the gas (using passive gas tracers) and are able to find a good match between the times of the detected $^{60}$Fe peaks and the inferred peak fluxes in their model.

A limitation of the Schulreich et al. \citep{Schulreich_etal_2017} model is that they do not consider sputtering of the grains along the way from the SN explosion to the Earth. They also ignore the magnetic field in the bubble. In our simulations, we find that the field morphology becomes quite complex and, although weak on average compared to its magnitude in the ambient medium, it has regions where it is strong enough to substantially deflect the grains. Most importantly, the assumption of close coupling of the grains to the gas is not well founded. 

In a project aimed at examining the ability of dust grains formed in SNe to escape into the ISM, we have carried out simulations that relax the assumption of close coupling (Slavin et al. 2020, ApJ submitted). Though there is evidence of efficient production of dust in SN ejecta, a longstanding question has been how much of that dust survives the passage of the reverse shock to escape into the ISM. 

Our simulations start with grains embedded in dense knots such as those observed in the Cas A SNR. Both the ejecta knots and the surrounding smooth ejecta are rapidly expanding into the surrounding medium. The expansion creates a forward shock and a reverse shock, which propagates back into the ejecta. The grains are subject drag and sputtering from the surrounding gas but are otherwise free to move independently of the gas. In this context the magnetic field is unlikely to be important and we do not include it.

\begin{SCfigure}[2.0][ht!]
\caption{Dust and gas for simulation of dust evolution in a young supernova remnant. The image shows the density (magnitude in cm$^{-3}$ is shown in the color bar) while the blue dots represent dust grains. It can be seen that most of the dust has propagated ahead of the expanding shock. Here the dust is assumed to be silicate grains that have an initial radius of $0.25$ $\mu$m. This is for a time 3000 yr after the explosion.}
\includegraphics[width=0.5\textwidth]{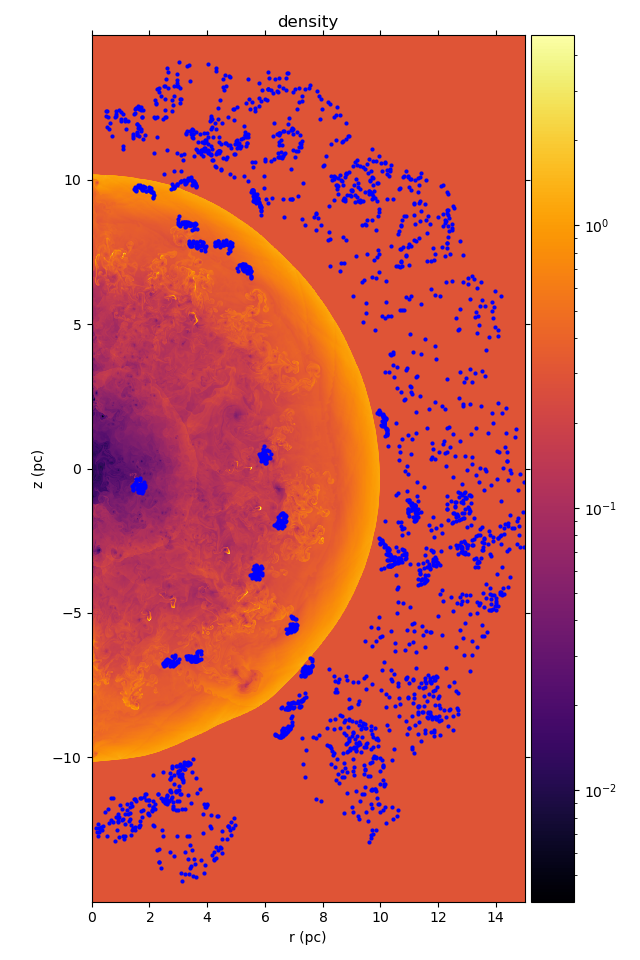}
\label{fig:CasA_dust}
\end{SCfigure}

As can be seen in Figure \ref{fig:CasA_dust} the grains fully decouple from the ejecta knots and can escape ahead of the forward shock. Here many of the dust grains (blue dots) have already escaped into the ISM ahead of the forward shock. However all have lost some fraction of their initial mass via sputtering and will lose more mass as they are slowed in the gas. We estimate that only roughly 10\% of the mass in these grains will remain after the grains have slowed below 20 km s$^{-1}$. However, larger grains can retain more of their initial mass. We speculate that grains formed in SNe that explode within pre-existing cavities may suffer less destruction, though such calculations have yet to be carried out.

\section{Capture of dust in the CLIC}\label{sect:capture}
The ability of SN created dust to escape the remnant raises interesting possibilities related to dust in the LIC. If large SN-created grains could have escaped from a nearby explosion and become embedded in the LIC, it could explain the surplus of large grains in the LIC -- grains added to the cloud without any addition of gas -- as well as the $^{60}$Fe dust seen in the Antarctic snow cores.

\begin{figure}[ht!]
    \centering
    \includegraphics[width=0.6\textwidth]{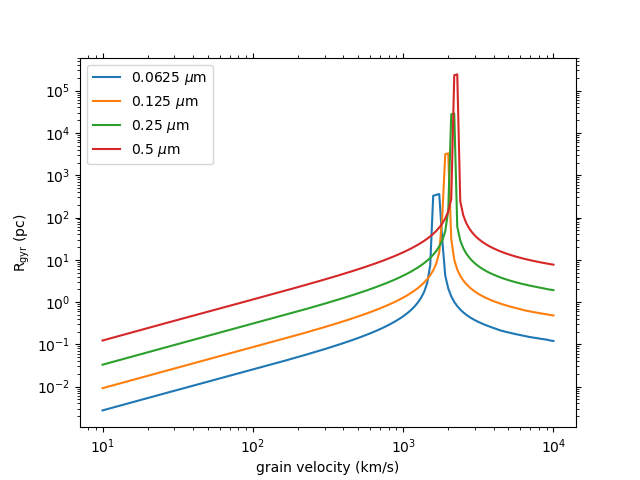}
    \caption{Gyroradius of silicate grains propagating through the a low density warm cloud such as the LIC with $B = 3\,\mu$G and $T = 7500\,$K.}
    \label{fig:gyroradii}
\end{figure}

However there are several stumbling blocks to this explanation of SN dust in the LIC. First, many models for SN dust formation find that most of the grains produced are small, $\lesssim 0.1\,\mu$m, though there are substantial uncertainties in these calculations. Second, the dust needs to be stopped without being destroyed. We discuss this further below.
Finally, the cloud magnetic field could simply reflect the grain back into the intercloud medium.

\begin{figure}[ht!]
    \centering
    \includegraphics[width=0.8\textwidth]{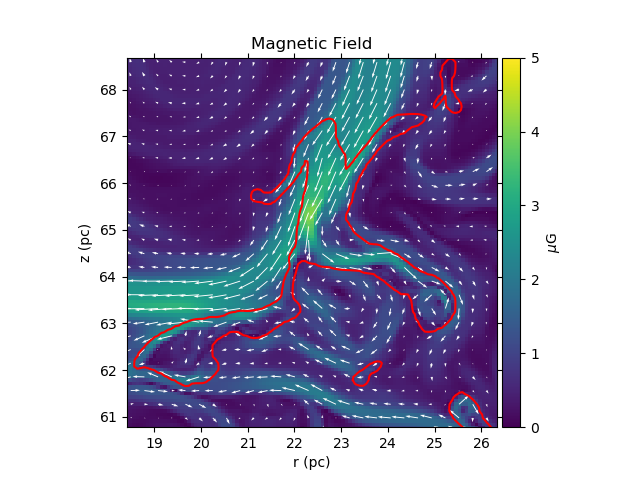}
    \caption{The magnetic field near a cloud from MHD simulations. The color corresponds to the magnitude of the field, while the arrows indicate the field direction. The red (density) contours show the cloud shapes.}
    \label{fig:B_field}
\end{figure}
Dust grains are charged in the ISM in general, by impact and sticking of ions and electrons and by photoelectric ejection of electrons. Their charge and mass lead to gyroradii in many situations that can be large compared even to relevant interstellar distance scales, e.g. the size of clouds, 
\begin{equation}
R_\mathrm{gyr} = \frac{8.5\, a(\mu m)^3\, v(\mathrm{km/s})}{|Z|\, B(\mu G)} \; \mathrm{pc},    \label{fig:rgyr}
\end{equation}
where $a$ is the grain radius, $v$ is the speed of the grain in the direction perpendicular to the field direction, $Z$ is the number of charges on the grain and $B$ is the the field strength. To find $Z$,
we have carried out calculations of the charging following \citep{McKee_etal_1987} and \citep{Weingartner+Draine_2001} for conditions like those in the LIC. These lead to gyroradii as shown in Figure \ref{fig:gyroradii}. Clearly for a fast moving grain, the gyroradius will exceed the size of the CLIC. More importantly, the magnetic field only diverts the path of the grains and does not slow them.

Our MHD calculations of the LISM that we mention in \S\ref{sect:LISM} show that the magnetic field in the Local Bubble is likely disordered at least if not fully turbulent, with the plasma $\beta \gg 1$ on average, though with small regions of larger field strength. We find that the field strength does not correlate simply with density as shown in Figure \ref{fig:B_field}. Such a disordered field could act to retain grains by deflecting them between clouds enough to allow them to slow. 

For the cloud to actually capture the grains they need to be slowed sufficiently that their velocity does not differ substantially from the gas they are embedded in. That can only be done by drag. The rule of thumb for grain slowing is that a grain needs to sweep up a mass of gas roughly equal to the grain mass. This can be expressed as requiring a column density of
\begin{equation}
    N \approx \frac{4}{3} a_{gr} \rho_{gr}/m \approx 1.7\times10^{20} a_{gr}(\mu m)\; \mathrm{cm}^{-2},
\end{equation}
where $\rho_{gr}$ is the solid material density of a grain and $m$ is the mean mass per particle of the gas and the numerical value assumes silicate grains. The LIC has a column density presently of only $\sim 10^{18}$ cm$^{-2}$ and even the entire CLIC has a total column density only a few times that.
\begin{figure}[ht!]
    \centering
    \includegraphics[width=0.7\textwidth]{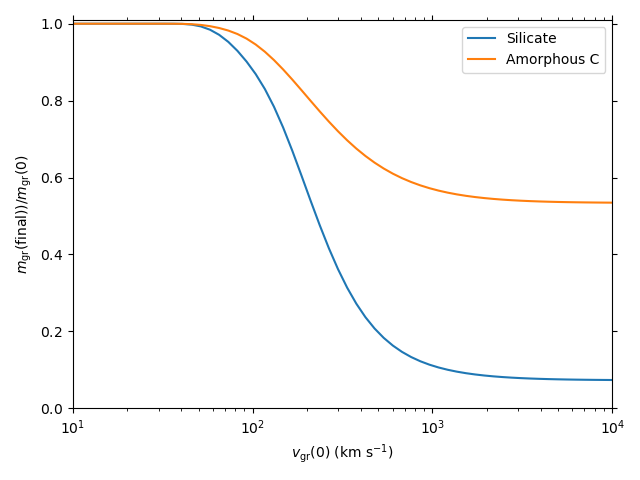}
    \caption{Fraction of initial mass remaining in a grain that is slowed by drag vs. the initial speed of the grain. This plot is for gas abundances like those in the LIC. Here the assumption is that grain is slowed to a speed below which it is no longer sputtered, which is $\sim 20$ km s$^{-1}$.}
    \label{fig:sputter}
\end{figure}
Another issue is that the same collisions with gas particles that slow grains sputter them as well. In fact it can be shown \citep{Micelotta_etal_2016} that the fraction of mass left in a grain as it is slowed (in the limit that only inertial sputtering is important, $T \lesssim 10^5$ K) is a only a function of the initial grain speed. This dependence is illustrated in Figure \ref{fig:sputter}. The assumed gas phase abundances are those for the LIC from \citep{Slavin+Frisch_2008} and the grains are assumed to have slowed below the threshold for sputtering, which for both silicate and carbonaceous grains is $\sim 20$ km s$^{-1}$. Thus to have slowed the largest interstellar grains observed in the heliosphere, $a \sim 1\,\mu$m, requires that they were initially larger than that by a substantial amount.

Given these pitfalls it would seem unlikely that the grains in the LIC were injected into the cloud from a nearby SN. However, there is one way out. In our simulations of the LISM, the clouds of the CLIC start out as cold, dense ($n = 28$ cm$^{-3}$) and smaller in volume as compared their current state. Their initial column density though, is substantially larger than that of the current LIC. We do not have any information about when the $^{60}$Fe dust got into the LIC, so it could have been a couple Myr ago and the cloud could have had a much larger column density at the time -- large enough to stop 1 $\mu$m grains.

\section{Summary}
\begin{itemize}
    \item The existence and characteristics of the Complex of Local Interstellar Clouds can be explained if they started as cold, neutral clouds embedded in warm ISM and were overrun by SN shocks that also heated the Local Bubble.
    \item The $^{60}$Fe in deep sea crusts is likely from dust created in one of the supernovae that created the Local Bubble, and reached Earth before it was in the LIC.
    \item The $^{60}$Fe in Antarctic snow can be explained if SN created dust was captured by the LIC, which may also explain the excess of large interstellar grains detected in the heliosphere.
    \item Explaining the $^{60}$Fe dust both in deep sea crusts and in Antarctic snow requires more careful examination of dust transport within the Local Bubble.
\end{itemize}

\ack We thank the organizers for the invitation to present this work and for putting on an excellent conference under trying circumstances. This research was supported by NASA Astrophysics Theory Program grant no.\ NNX17AH80G and a Smithsonian Institution Scholarly Studies grant. We thank Priscilla Frisch for helpful conversations. We are grateful to Gunther Korschinek, Thomas Faestermann and Joe Weingartner for permission to use their figures.

\bibliographystyle{iopart-num}
\bibliography{lism}

\end{document}